\magnification=\magstep1
\vsize=20truecm
\hsize=14truecm
\voffset=2.5truecm
\hoffset=1.0truecm
\baselineskip=0.5truecm
\parskip=0.5truecm
\parindent=0.0truecm
\font\title=cmbx10 scaled 2000
\font\author=cmti10 scaled 1500
\font\address=cmr10 scaled 1250
\font\date=cmr10 scaled 1125
\font\report=cmr10 scaled 1063

{\title
\centerline{MY LIFE AS TUTOR}
\vskip 1.5\baselineskip
\centerline{Reflections on Two Recent Experiences}}
\vskip 5.0\baselineskip
{\author
\centerline{Alessandro B. Romeo}}
\vskip 1.5\baselineskip
{\address
\centerline{Onsala Space Observatory,}
\vskip 0.5\baselineskip
\centerline{Chalmers University of Technology,}
\vskip 0.5\baselineskip
\centerline{S-43992 Onsala, Sweden}
\vskip 0.5\baselineskip
\centerline{(romeo@oso.chalmers.se)}}
\vskip 5.0\baselineskip
{\date
\centerline{22 January 1999}}
\vskip 5.0\baselineskip
{\report
\centerline{Final Report for `Pedagogy of Group Tutoring' (1998),}
\vskip 0.5\baselineskip
\centerline{Advanced Research Course at Chalmers University of Technology}}
\vfill\eject

{\bf SUMMARY}

In this final report, I briefly reflect on two parallel teaching experiences
as tutor: one for `Computer Science and Engineering in Context' (1998) and
the other for `Introduction to Electrical Engineering' (1998), obligatory
courses at Chalmers University of Technology.  Special emphasis is put on the
former.  Besides, I briefly view such experiences in interaction with my
research work, private life and new teaching position.  In harmony with my
conception of teaching, I avoid the standard formal style of reports and try
an interactive dialogue with the reader.

\vfill\eject

{\bf Why I am writing this report}

The course `Computer Science and Engineering in Context' is finished and now
it is we, the tutors, who should write a final report on our teaching
experience.  Such reflections are important for at least two reasons: (1) for
ourselves, because we can view our experience retrospectively and learn how
to improve our role of educators; (2) for the others, either new tutors or
course organizers, in order to improve the structure of the course and
optimize its impact on the student education at Chalmers.

In writing this report, I will use a colloquial style and imagine to interact
directly with the reader.  I always use a formal style in writing scientific
papers.  Here instead I want to experiment with a new way of communicating my
ideas, something between an oral discussion and a written report, something
more spontaneous and hopefully more effective.  So let us discuss the main
phases of my experience.

\vskip 2\baselineskip

{\bf How I got involved in this course}

I had just come back from a three-month visit to the International School for
Advanced Studies in Trieste, Italy, when I heard that Peter Jansson was
recruiting tutors for the course.  At that time, I had an urgent need of
financial support since my position at Onsala Space Observatory was finished
a few months before.  So I didn't hesitate to confirm my participation.  On
the other hand, I did also hope to complete the scientific work started in
Italy, and thus I was unwilling to teach intensively before the new year.

The first meeting with Peter and the other tutors was a really positive
start.  What I liked most of that workshop was, apart from the beautiful
place where it was held, the fact that it was structured as an interactive
discussion between course organizers, pedagogues and tutors.  This is
precisely what I believe teaching should be: not a monologue, but an open
dialogue between the participants.  My previous teaching experience taught me
that flexibility and intuition are important qualities in such a respect.
After this first workshop, I felt very charged and ready to start.

\vskip 2\baselineskip

{\bf What I expected and what happened}

Before the actual start of the course I felt somewhat anxious.  In fact, this
was the first time that I should tutor research projects in other fields than
physics or mathematics.  I am an astrophysicist and know very little about
computer science and engineering.  Students always expect much from their
teachers, and so I expected much from myself.  But how could I explain them
that I don't know anything about their field and, yet, I could be a good
tutor?  The best would be of course to clarify my role since the beginning,
and so I did.

The initial expectation analyses, one performed on myself and the other
performed on my students, gave immediately good results.  I felt accepted by
the students, and thus I could also trust myself in the new role.  This was
the premise for a fruitful atmosphere and a successful work.

\vskip 2\baselineskip

{\bf How I acted: an ABC}

{\bf A.}  Next came the problem of how to integrate the important points
emphasized at the pedagogy course into my personal method of teaching.  My
conception of teaching is that it should be dynamic, as opposed to static.
So what better opportunity than experimenting brainstorming, snowball and
related techniques since the beginning, that is at our first group meeting?
Students accepted this idea enthusiastically.  They like to experiment new
forms of learning, just as we like to experiment new forms of teaching.  This
is because our mind likes changes.  Nobody can guarantee that one method is
better than another, it depends on so many factors, most of which are out of
our control.  What is important is to adapt ourselves as teachers to the
students, but such a process should be as swift and smooth as possible.
Besides, perhaps even more important, we should show the students that we are
willing to learn from them.  I believe that such a conception of reciprocal
teaching and learning is what can make teachers and students active members
of the same team: to give and to receive work always better together.

{\bf B.}  Now a crucial question arises: how can we optimize our interaction
with the students?  The answer that I found is: observing how students
interact between one another, and deciding which role to take accordingly.
If we want to observe the real behaviour of students, we should disturb them
as little as possible with our presence and make them unaware of our role of
observers.  My solution was to say: ``I see that you are discussing
intensively specific topics of the project.  Do you mind if, meanwhile, I
read something urgent?''  So, while pretending to read an astrophysical
paper, I was instead concentrating on their interactions.  How long?  About
half an hour, that is the typical time that students need for forgetting
about the presence of an apparently inactive tutor.  How many times?  About
once or twice a month, in order to monitor the evolution of the student
interactions.  This is especially important for finding out if there are
conflicts in the group.  In such a case, the best is of course to sort out
these conflicts as soon as possible, and the tutor should then act as a real
psycho-analyst.

{\bf C.}  Last but not least, what about student evaluation?  A key factor to
bear in mind is that, whenever people interact, they evaluate one another.
This means that not only teachers evaluate students, but also students
tacitly evaluate teachers.  Both evaluations start from the beginning and
last up to the end of the series of meetings, or even beyond.  Our goal as
educators is to give students good examples even concerning such a delicate
point: we should show them that timely and constructive evaluations can be of
benefit for both parts.  We are human beings, and mistakes belong to our
nature.  It is through mistakes that children learn to live in the adults'
world, it is through mistakes that humanity has progressed to the present.
The important thing is to recognize our mistakes and to improve.  Yes, if we
emphasize this goal as the reason of the evaluations, then we can be sure
that students will agree and collaborate.  What a better example than to
start with ourselves?  It is hard but necessary.  If we show that we want to
improve through the suggestions of the students, then we also show that we
are willing to learn from them.  What a better satisfaction for the students?
To teach their teachers!  Everyone teaches and learns at the same time and
democratically, as it should be.

{\bf In summary:} Be a good example for your students not only as an
educator, but also as a student and a human being!

\vskip 2\baselineskip

{\bf If you want to know more}

The previous discussion can be extended by referring to my four progress
reports, which are unpublished but available on request, and to appropriate
literature.  Let me first recommend the book and TV series: `Filosofiska
Fr{\aa}gor -- \"Aventyr i Tankens V\"arld' produced by the
`UtbildningsRadion' in 1998.  Such a source does not directly concern
pedagogy, but it helps us to reflect on fundamental problems about ourselves
and our complex interactions with others, and thus to learn from our life
experiences.  The literature suggested at the pedagogy course covers
important aspects of group tutoring and related problems.  Lenn\'eer-Axelson
\& Thylefors (1991) discusses the psycho-sociology of group work, and I find
it especially useful for learning how to cope with conflicts.  Jaques (1992)
discusses educational methods for improving the quality of group teaching and
training.  I find it especially useful for learning several things:
techniques such as brainstorming and snowball (expectation analysis is
discussed by Widerberg 1994); the tutor's roles in his/her interaction with
the students, and their effects on group dynamics and evolution; and group
evaluation.  In addition, Booth et al.\ (1997) and Jansson (1998) analyse the
impacts of the course `Computer Science and Enginnering in Context' and the
underlying project `D++' on the Swedish education.

\vskip 2\baselineskip

{\bf What I learned}

What I learned from this experience is due to two major sources: (1) the
excellent course `Pedagogy of Group Tutoring' organized by Peter Jansson and
Shirley Booth; (2) the interaction with my students.  The first source
represented a concrete example of how fruitful a group work can be.  It also
contributed to structure and rationalize my ideas about teaching, which had
started to take form about 15 years ago, that is at the time of my first
private lessons.  The second source was, I believe, fundamental: I learned to
learn from my students.

\vskip 2\baselineskip

{\bf Comparison with a parallel experience}

My experience as tutor for the course `Computer Science and Engineering in
Context' was rich and joyful.  I had an ideal group of students, the best
among all first-year students that I had had in five years of teaching here
at Chalmers.  The course organization was excellent, a real team work between
course organizers, pedagogues and tutors.  What else to say?

Parallel to that, I had a painful experience as tutor for the course
`Introduction to Electrical Engineering'.  The group of students was on the
whole very lazy.  Actually, there was a good student among them, but he was
shy.  So the group was dominated by the lack of internal motivation and the
stubbornness of the other students.  I did my best to stimulate their
enthusiasm and activity, but with poor results.  The course organization was,
to say the least, chaotic.  The recruitment of tutors was completed more than
one week after the start of the course.  Tutors were not informed about their
duties until few days before they were assumed to act.  For example, they had
to play the role of mathematical `exercisers' (not tutors) with no previous
psychological preparation for that, and with a resulting substantial delay in
the start of the project.  All the steps of the project were decided by the
course organizer without consulting the tutors, who just found weekly
instructions for their work .......... The list is even longer
.......... Last, and worst, I had a clear impression that the course
organizer wanted to convey the idea that students should pass the exam; this
impression was also shared by the students of my group.  What else to say?
An experience that I would never repeat!

One may wonder whether such a picture is objective or not.  How is it
possible that two parallel experiences were one so positive and the other so
negative?  What I can say is that I am not black and white in the way of
thinking, and that I have other teaching experiences for comparison.  Thus I
believe that the picture reflects sharply the contrast between two course
organizations, which unavoidably influence the performance of both tutors and
students.  Viewing such organizations in context and monitoring the opinions
of both tutors and students may help these courses to progress towards their
important common goal.

\vskip 2\baselineskip

{\bf How teaching interacted with my research work}

When I came back from Italy, I was very excited and full of new ideas.  It
was the first time that I could stay in my natural environment for more than
one month after seven years of residence in Sweden.  At the International
School for Advanced Studies I had started three important scientific
collaborations, one of which together with the Astrophysics Group at
Chalmers, and I had planned to complete these projects as soon as possible
because of the official end of my position at Onsala Space Observatory.  Thus
the impact of intensive teaching on my research work was hard at the
beginning: I was obliged to teach in order to earn money for surviving.  This
was in sharp contrast to my previous teaching experience, in which I had got
involved spontaneously.  Did anything change afterwards?  Yes, definitely!
Peter and my students of the course `Computer Science and Engineering in
Context' succeeded to create such a stimulating atmosphere around my role of
tutor that I quickly became enthusiastic, even considering my disappointment
about the other course.  (Now, of course, I feel a bit embarrassed to write
such things about Peter because he will read my report, but this is the pure
truth.)  In conclusion, my research work underwent substantial delays, but I
earned an invaluably rich teaching experience.

\vskip 2\baselineskip

{\bf How teaching interacted with my private life}

Outside the academic world, I am a happy husband and father of three
children.  Apart from an obvious period of stress caused by the initial
teaching-research conflict, my family life reflected the satisfaction of
being tutor.  I was so absorbed in such a role that I proposed a practical
project for the week-ends to my family.  A brief work plan was ready by the
end of November, and the result was the construction of a beautiful Christmas
crib.  There was no final report :-)

\vfill\eject

{\bf MY NEW LIFE AS LECTOR}

Now my children would say: ``snipp snapp snut, nu \"ar sagan slut'', but the
end of that story is the beginning of a new one.  This Christmas I have
received a surprising present: a tenure-track lector position, which the
Swedish Natural Science Research Council will financially support to 80\% for
three years, that is beyond the start of the new millennium.  This means a
lot to me: more balance between my teaching and research works, possibly time
for popular science, more economical security for my family, and thus a
clearer view of my future; in other words, a new life.


\bye